\documentclass[11pt]{article}

\usepackage{amsmath,amssymb,amsfonts,bm}
\usepackage{graphicx}
\usepackage{subcaption}
\usepackage{times}
\usepackage{url}
\usepackage{xcolor}
\usepackage[colorlinks=true,
            urlcolor=blue,
            linkcolor=black,
            citecolor=black]{hyperref}

\usepackage{geometry}
\begin{document}
\sloppy

\vspace{-15pt}
\begin{flushleft}

\huge{\bf
Directional reflection modeling via wavenumber-domain reflection coefficient
for 3D acoustic field simulation
}

\vspace{6pt}

\large{\bf Satoshi Hoshika and Takahiro Iwami}\par
\normalsize{\it
Graduate School of Design, Kyushu University, Fukuoka, Japan\\
hoshikasatoshi@gmail.com,\; iwami@design.kyushu-u.ac.jp
}

\vspace{6pt}

\large{\bf Akira Omoto}\par
\normalsize{\it
Faculty of Design, Kyushu University, Fukuoka, Japan\\
omoto@design.kyushu-u.ac.jp
}

\end{flushleft}

\begin{abstract}
\noindent
{ \normalsize
This study proposes a framework for incorporating wavenumber-domain acoustic
reflection coefficients into sound field analysis to characterize
direction-dependent material reflection and scattering phenomena. The
reflection coefficient is defined as the amplitude ratio between incident and
reflected waves for each propagation direction and is estimated from spatial
Fourier transforms of the incident and reflected sound fields. The resulting
wavenumber-domain reflection coefficients are converted into an acoustic
admittance representation that is directly compatible with numerical methods
such as the Boundary Element Method (BEM), enabling simulation of reflections
beyond simple specular components.
Unlike conventional extended reaction models, the proposed approach avoids
explicit modeling of the material interior. This significantly reduces
computational cost while allowing direct use of measured data, empirical
models, or user-defined directional reflection characteristics.
The validity of the proposed formulation was previously demonstrated by the
authors through two-dimensional sound field simulations, in which accurate
reproduction of direction-dependent reflection behavior was confirmed.
In the present work, the framework is extended to three-dimensional analysis,
demonstrating its applicability to more realistic and complex acoustic
environments.
The proposed approach provides a practical and flexible tool for simulating
direction-dependent acoustic reflections and scattering, with potential
applications in architectural acoustics, material characterization, and noise
control.
}
\end{abstract}

\noindent
\textit{Submitted to Proceedings of Meetings on Acoustics (PoMA 2025).}


\section{Background and Motivation}

Impedance boundary conditions have long been used as a standard approach for
modeling sound absorption and reflection at material surfaces.
In their simplest form, such boundary conditions relate the normal particle
velocity to the sound pressure at the surface and provide an efficient means of
representing energy dissipation without explicitly resolving the interior of the
material.
However, locally reactive impedance models inherently neglect directional
effects and are unable to describe non-specular reflection or scattering
phenomena.

To address these limitations, extended reaction models have been developed to
account for wave propagation and energy dissipation within the material
interior.
Representative examples include equivalent fluid models
\cite{Johnson1987,Champoux1991}, Biot-type poroelastic models
\cite{Biot1956a,Biot1956b}, and numerical analyses of periodic or layered
structures \cite{Kushwaha1993,Groby2011}.
While these approaches enable the prediction of angle-dependent reflection and
scattering characteristics, they require volumetric discretization of the
material domain.
In three-dimensional simulations, this leads to a substantial increase in
computational cost, often resulting in meshes consisting of several tens or
hundreds of thousands of elements.
Furthermore, resolving directional reflection behavior typically requires
multiple simulations for different incident angles, which further compounds the
computational burden.

In contrast to these approaches, the present study adopts a macroscopic
viewpoint that focuses exclusively on the relationship between incident and
reflected waves at the boundary.
Rather than modeling the internal dynamics of the material, the proposed
framework treats the boundary as a linear operator that redistributes incident
acoustic energy among reflected directions.
This relationship is described in terms of propagation directions by applying a
spatial Fourier transform to the boundary pressure field, which enables the
incident and reflected sound fields to be decomposed into plane-wave components
indexed by in-plane wavenumbers.
This perspective is particularly suited to applications where the internal
microstructure is unknown, too complex to model explicitly, or unnecessary for
describing the observable reflection behavior.

Spatial Fourier transform techniques have previously been employed to analyze
directional reflection characteristics at boundaries.
For example, Tamura proposed a method to estimate angle-dependent reflection
coefficients from measured sound pressure distributions using a spatial Fourier
transform \cite{tamura1,tamura2}.
Although this approach enables the evaluation of oblique incidence effects, the
reflection coefficient is treated as a scalar quantity defined independently for
each incident direction.
As a result, non-specular reflection and directional coupling between different
wave components cannot be represented within a unified framework.

To overcome this limitation, the authors previously introduced the concept of a
\emph{wavenumber-domain acoustic reflection coefficient}, denoted by
$\mathbf{C}_{\mathrm{r}}$, which represents a linear mapping between incident and
reflected wave components in wavenumber space
\cite{Hoshika2025,Hoshika_IN2024}.
This operator-based formulation naturally captures non-specular reflection and
multi-directional scattering without invoking volumetric models of the material
interior.
However, previous studies were primarily limited to two-dimensional sound
fields, and a systematic extension to three-dimensional problems, as well as a
practical integration of the estimated reflection characteristics into
numerical sound field simulations, had not yet been established.

The objective of the present study is to construct a unified framework for
describing directional acoustic reflection in three-dimensional sound fields.
The proposed method estimates the wavenumber-domain reflection matrix
$\mathbf{C}_{\mathrm{r}}$ from spatial distributions of incident and reflected
waves obtained using multiple sound sources and receivers.
The estimated matrix is then transformed into a wavenumber-domain admittance and
incorporated into a conventional boundary element method (BEM) as a non-local
boundary condition.
This enables complex angle-dependent reflection and scattering behavior to be
reproduced directly at the boundary, while avoiding the high computational cost
associated with extended reaction models that explicitly resolve the material
interior.

\section{Principle of the Wavenumber-Domain Reflection Coefficient Estimation in 3D Field}
\subsection{Definition of the Wavenumber-Domain Acoustic Reflection Coefficient}
\label{sec:def_cr}

\begin{figure}[t]
\centering
\includegraphics[width=0.6\columnwidth]{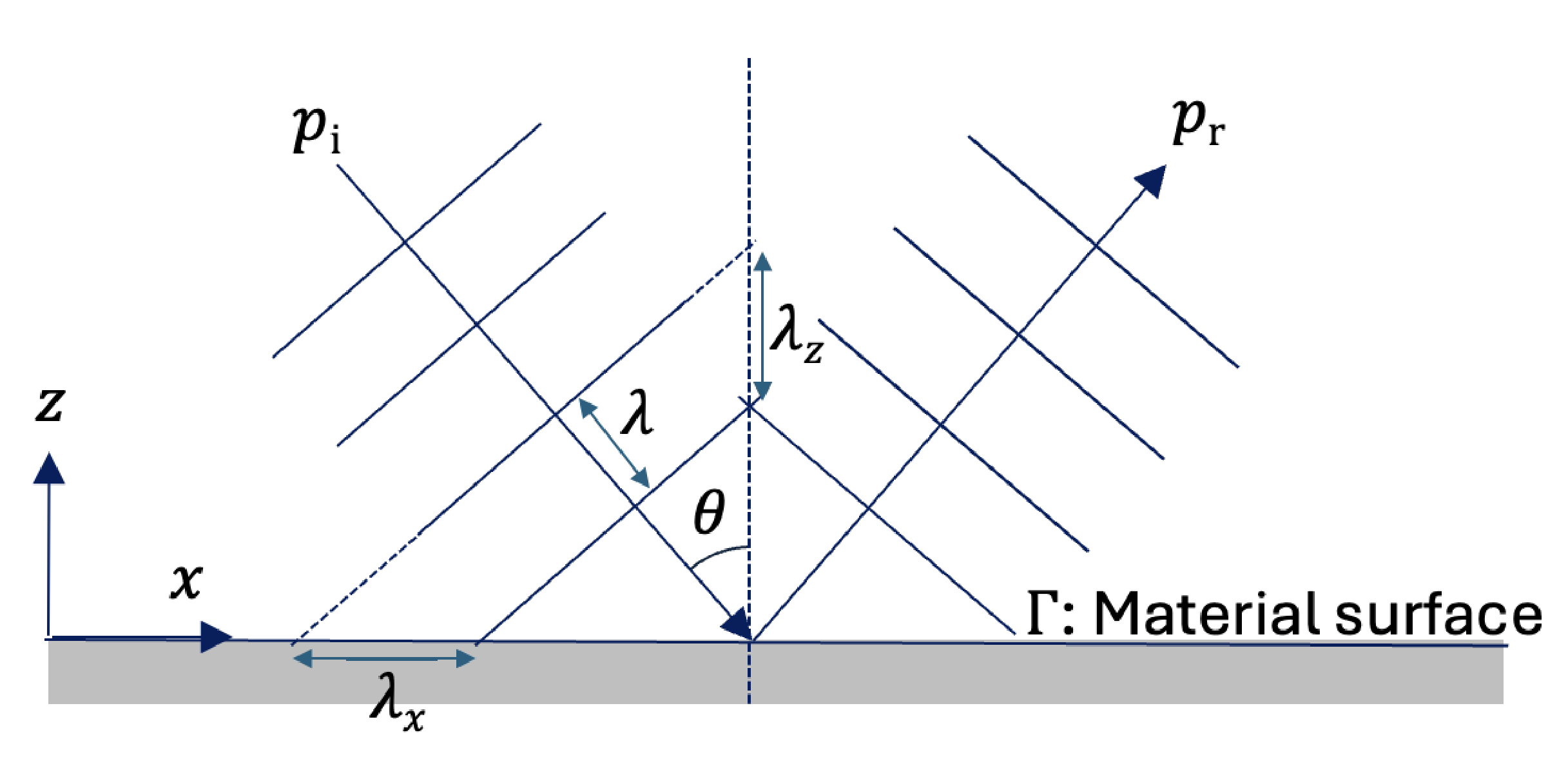}
\vspace{-0pt}
\caption{Schematic illustration of reflection in a three-dimensional acoustic field.
A plane wave is incident from the upper half-space onto the boundary surface at $z=0$,
where it is reflected back into the same half-space.}
\label{fig:3d_field1}
\vspace{-0pt}
\end{figure}

We consider an incident and a reflected acoustic wave in a three-dimensional
acoustic field, as illustrated in Fig.~\ref{fig:3d_field1}.
The material surface is located at
\(
\Gamma = \{\, \bm{r}=(x,y,z)\in\mathbb{R}^3 \mid z = 0 \,\},
\)
and the incident and reflected sound pressures are denoted by
\(p_{\mathrm{i}}\) and \(p_{\mathrm{r}}\), respectively.
All quantities are treated in the frequency domain, and the angular frequency
\(\omega\) is omitted for brevity.

We regard the pressure distribution on the boundary plane and its Fourier
transform as finite-energy signals and adopt the Hilbert space
\(L^2(\mathbb{R}^2)\) as the functional setting.
In wavenumber-domain analysis, only the components parallel to the boundary,
\((k_x,k_y)\), are considered:
\[
\bm{k}_{xy} := (k_x,k_y), \qquad
\mathcal{D}_k := \{\, \bm{k}_{xy}\in\mathbb{R}^2 \mid k_x^2 + k_y^2 \le k^2 \,\}.
\]
Evanescent components satisfying \(|\bm{k}_{xy}|>k\) do not contribute to the
radiated far field and are therefore neglected.

The two-dimensional Fourier transform along the boundary plane is defined as
\cite{williams}
\begin{align}
\mathcal{F}_{xy} f(\bm{k}_{xy})
&:= \frac{1}{2\pi}\iint_{\mathbb{R}^2}
f(x,y)\, e^{-i(k_x x + k_y y)}\, \mathrm{d}x\,\mathrm{d}y,
\label{eq:fourier2d_def} \\
\mathcal{F}_{xy}^{-1}\hat{f}(x,y)
&:= \frac{1}{2\pi}\iint_{\mathbb{R}^2}
\hat{f}(\bm{k}_{xy})\, e^{i(k_x x + k_y y)}\,
\mathrm{d}k_x\,\mathrm{d}k_y .
\label{eq:fourier2d_inv}
\end{align}

Representing the three-dimensional wavenumber vector in spherical coordinates,
\begin{align}
k_x &= k \sin\theta \cos\phi, \qquad
k_y = k \sin\theta \sin\phi, \qquad
k_z = k \cos\theta ,
\label{eq:3d_wavevector}
\end{align}
where for propagating components
\(
|k_z| = \sqrt{k^2 - k_x^2 - k_y^2}.
\)

Using this representation, the incident and reflected wavenumber-domain spectra
are defined on the same domain $\mathcal{D}_k$ but are distinguished by the sign
of the normal wavenumber component:
\begin{align}
\hat{p}_{\mathrm{i}}(\bm{k}_{xy})
&:= \mathcal{F}_{xy} p_{\mathrm{i}}(\bm{k}_{xy}),
\qquad k_z < 0, \\
\hat{p}_{\mathrm{r}}(\bm{k}_{xy})
&:= \mathcal{F}_{xy} p_{\mathrm{r}}(\bm{k}_{xy}),
\qquad k_z > 0 .
\end{align}
That is, $\hat{p}_{\mathrm{i}}$ represents plane-wave components propagating
toward the boundary, whereas $\hat{p}_{\mathrm{r}}$ corresponds to components
propagating away from the boundary after reflection.

The incident and reflected fields can then be expressed by plane-wave
expansions as
\begin{align}
p_{\mathrm{i}}(x,y,z)
&= \frac{1}{2\pi}\iint_{\mathcal{D}_k}
\hat{p}_{\mathrm{i}}(\bm{k}_{xy})\,
\mathrm{e}^{i(k_x x + k_y y - |k_z| z)}\,
\mathrm{d}\bm{k}_{xy},
\label{eq:pw3d_inc} \\
p_{\mathrm{r}}(x,y,z)
&= \frac{1}{2\pi}\iint_{\mathcal{D}_k}
\hat{p}_{\mathrm{r}}(\bm{k}_{xy})\,
\mathrm{e}^{i(k_x x + k_y y + |k_z| z)}\,
\mathrm{d}\bm{k}_{xy}.
\label{eq:pw3d_ref}
\end{align}

Using the boundary-plane spectra, the reflection operator can be written as
\begin{align}
\hat{p}_{\mathrm{r}}(\bm{k}_{xy})
=
\iint_{\mathcal{D}_k}
C_{\mathrm{r}}(\bm{k}_{xy}, \bm{k}'_{xy})\, \hat{p}_{\mathrm{i}}(\bm{k}'_{xy})\,
\mathrm{d}\bm{k}'_{xy},
\label{eq:Cr_operator3d_full}
\end{align}
where the reflection operator
\(
\mathcal{C}_{\mathrm{r}}: L^2(\mathcal{D}_k) \to L^2(\mathcal{D}_k)
\)
is defined as
\[
\mathcal{C}_{\mathrm{r}} \hat{f}(\bm{k}_{xy})
:=
\iint_{\mathcal{D}_k}
C_{\mathrm{r}}(\bm{k}_{xy}, \bm{k}'_{xy})\, \hat{f}(\bm{k}'_{xy})\,
\mathrm{d}\bm{k}'_{xy}.
\]
Considering energy balance and dissipation, the operator norm satisfies
\(
\lVert \mathcal{C}_{\mathrm{r}} \rVert_{\mathrm{op}} \le 1
\),
and strictly
\(
\lVert \mathcal{C}_{\mathrm{r}} \rVert_{\mathrm{op}} < 1
\)
unless the boundary is perfectly reflecting \cite{MorseIngard1986}.

For numerical implementation, the wavenumber domain $\mathcal{D}_k$ is
discretized by sampling points
\(
\{\bm{k}_{xy,m}\}_{m=1}^{M}
\)
with associated quadrature weights
\(
w_m > 0
\),
which represent the integration measure corresponding to the finite area in
the wavenumber plane covered by each discrete mode.
Accordingly, the continuous integral is approximated as
\[
\iint_{\mathcal{D}_k} f(\bm{k}_{xy})\,\mathrm{d}\bm{k}_{xy}
\;\approx\;
\sum_{n=1}^{M} f(\bm{k}_{xy,n})\, w_n .
\]
The reflected spectrum is thus discretized as
\[
\hat{p}_{\mathrm{r}}(\bm{k}_{xy,m})
\approx
\sum_{n=1}^{M}
C_{\mathrm{r}}(\bm{k}_{xy,m}, \bm{k}_{xy,n})\,
\hat{p}_{\mathrm{i}}(\bm{k}_{xy,n})\, w_n .
\]

The specific form of $w_n$ depends on the sampling scheme used to discretize
$\mathcal{D}_k$.
For example, when a Fibonacci lattice is employed to uniformly sample the upper
hemisphere, the quadrature weight can be approximated as
\[
w_n \approx \frac{2\pi k k_{z,n}}{M},
\]
where $k$ is the acoustic wavenumber, $k_{z,n}$ is the normal component of the
$n$-th wavenumber vector, and $M$ is the total number of sampled modes.
This choice ensures an approximately uniform partition of solid angle and
energy-consistent discretization of the reflection operator.

\noindent Define the vectors
\begin{align}
\hat{\mathbf{p}}_{\mathrm{i}}
&=
[\hat{p}_{\mathrm{i}}(\bm{k}_{xy,1}), \dots,
 \hat{p}_{\mathrm{i}}(\bm{k}_{xy,M})]^{\mathsf{T}}, \\
\hat{\mathbf{p}}_{\mathrm{r}}
&=
[\hat{p}_{\mathrm{r}}(\bm{k}_{xy,1}), \dots,
 \hat{p}_{\mathrm{r}}(\bm{k}_{xy,M})]^{\mathsf{T}} .
\end{align}

\noindent Defining the matrix
\[
[\mathbf{C}_{\mathrm{r}}]_{mn}
:= C_{\mathrm{r}}(\bm{k}_{xy,m}, \bm{k}_{xy,n})\, w_n ,
\]
the discrete relation can be written as
\begin{align}
\hat{\mathbf{p}}_{\mathrm{r}}
=
\mathbf{C}_{\mathrm{r}}\, \hat{\mathbf{p}}_{\mathrm{i}} .
\label{eq:Cr_matrix3d}
\end{align}

Diagonal elements of $\mathbf{C}_{\mathrm{r}}$ correspond to specular reflection,
whereas off-diagonal elements represent scattering-induced directional coupling.

\subsection{Estimation of the Wavenumber-Domain Reflection Coefficient (3D)}

Based on Eq.~\eqref{eq:Cr_matrix3d}, the wavenumber-domain reflection matrix
$\mathbf{C}_{\mathrm{r}}$ is estimated from $Q$ sets of incident and reflected
spectra.
Stacking these spectra column-wise yields
\begin{align}
\hat{\bm{\mathcal{P}}}_{\mathrm{i}}
&=
[\hat{\mathbf{p}}_{\mathrm{i},1}, \dots,
 \hat{\mathbf{p}}_{\mathrm{i},Q}]
 \in \mathbb{C}^{M\times Q}, \\
\hat{\bm{\mathcal{P}}}_{\mathrm{r}}
&=
[\hat{\mathbf{p}}_{\mathrm{r},1}, \dots,
 \hat{\mathbf{p}}_{\mathrm{r},Q}]
 \in \mathbb{C}^{M\times Q}.
\end{align}
These matrices satisfy the linear relation
\begin{align}
\hat{\bm{\mathcal{P}}}_{\mathrm{r}}
&=
\mathbf{C}_{\mathrm{r}}
\hat{\bm{\mathcal{P}}}_{\mathrm{i}},
\end{align}
from which $\mathbf{C}_{\mathrm{r}}$ can be estimated, for example, using the
Moore--Penrose pseudoinverse as
\begin{align}
\mathbf{C}_{\mathrm{r}}
\simeq
\hat{\bm{\mathcal{P}}}_{\mathrm{r}}
\hat{\bm{\mathcal{P}}}_{\mathrm{i}}^{\dag},
\end{align}
where $(\cdot)^{\dag}$ denotes the pseudoinverse.

For boundaries with homogeneous or periodic structure, the reflected wavefield
often concentrates into a limited number of wavenumber components.
As a result, the corresponding reflection matrix $\mathbf{C}_{\mathrm{r}}$
tends to exhibit a sparse or approximately sparse structure in wavenumber space.
When such structure is anticipated, sparsity-promoting estimation techniques,
such as $\ell_1$-regularized regression, can be employed as an alternative
solution strategy to enhance robustness under limited or noisy observations.
In the present study, this type of regularization is used as a practical option,
while the proposed framework itself is not restricted to any specific estimation
method.


\section{Admittance Operator and Its Application to Numerical Simulations}
\subsection{Derivation of the Normalized Acoustic Admittance Operator and Its Matrix Representation}

As shown in the previous section, the reflection coefficient in the wavenumber domain
describes how an incident wave is reflected at a boundary in terms of both direction
and amplitude.
In this section, we present a framework for applying this theory to numerical simulations
by transforming the reflection coefficient into a normalized acoustic admittance.
This formulation is particularly useful for numerical methods in which boundary
conditions are expressed in terms of admittance, such as the boundary element method (BEM).
To this end, the admittance operator is derived from the relationship between sound
pressure and particle velocity in the wavenumber domain.

We define the normalized acoustic admittance operator
\((\mathcal{B} : L^2(\mathcal{D}_k) \to L^2(\mathcal{D}_k))\),
which maps sound pressure to particle velocity in the wavenumber domain, as
\begin{equation}
\mathcal{B}\, \hat{p}(\bm{k}_{xy}) = \hat{v}(\bm{k}_{xy}).
\label{eq:admittance}
\end{equation}

The sound pressure field in the wavenumber domain,
\(\hat{p}(\bm{k}_{xy}, z)\),
can be expressed as the superposition of the incident and reflected waves as
\begin{equation}
\hat{p}(\bm{k}_{xy}, z)
= \hat{p}_{\mathrm{i}}(\bm{k}_{xy}) \mathrm{e}^{-i k_z z}
+ \hat{p}_{\mathrm{r}}(\bm{k}_{xy}) \mathrm{e}^{i k_z z}.
\end{equation}

Let \(\rho\) and \(c\) denote the density and sound speed of the medium, respectively.
The particle velocity in the \(z\)-direction,
\(\hat{v}(\bm{k}_{xy}, z)\),
is obtained from Euler's equation as
\begin{align}
\hat{v}(\bm{k}_{xy}, z)
= \frac{1}{i \rho \omega}
\frac{\partial}{\partial z}
\hat{p}(\bm{k}_{xy}, z) 
= \frac{k_z}{\rho c k}
\left(
- \hat{p}_{\mathrm{i}}(\bm{k}_{xy}) \mathrm{e}^{-i k_z z}
+ \hat{p}_{\mathrm{r}}(\bm{k}_{xy}) \mathrm{e}^{i k_z z}
\right).
\end{align}

At the boundary \(z = 0\), the sound pressure and particle velocity are given by
\begin{align}
\hat{p}(\bm{k}_{xy})
&= \hat{p}_{\mathrm{i}}(\bm{k}_{xy})
+ \hat{p}_{\mathrm{r}}(\bm{k}_{xy}), \\
\hat{v}(\bm{k}_{xy})
&= \frac{k_z}{\rho c k}
\left(
- \hat{p}_{\mathrm{i}}(\bm{k}_{xy})
+ \hat{p}_{\mathrm{r}}(\bm{k}_{xy})
\right).
\end{align}

Using the reflection coefficient operator \(\mathcal{C}_r\) and the identity operator
\(\mathcal{I}\), the pressure and particle velocity can be rewritten as
\begin{align}
\hat{p}(\bm{k}_{xy})
&= (\mathcal{I} + \mathcal{C}_r)\,
\hat{p}_{\mathrm{i}}(\bm{k}_{xy}),
\label{eq:P_kxy} \\
\hat{v}(\bm{k}_{xy})
&= \hat{\beta}_0(\bm{k}_{xy})
(-\mathcal{I} + \mathcal{C}_r)\,
\hat{p}_{\mathrm{i}}(\bm{k}_{xy}),
\label{eq:V_kxy}
\end{align}
where \(\hat{\beta}_0(\bm{k}_{xy})\) denotes the characteristic admittance corresponding
to the in-plane wavenumber \(\bm{k}_{xy}\), defined as
\begin{equation}
\hat{\beta}_0(\bm{k}_{xy})
:= \frac{k_z}{\rho c k}
= \frac{\sqrt{k^2 - (k_x^2 + k_y^2)}}{\rho c k}.
\end{equation}
This quantity is equal to
\(\hat{\beta}_0(\bm{k}_{xy}) = \cos \theta / (\rho c)\),
where \(\theta\) is the angle of incidence, and represents the characteristic admittance
for obliquely incident waves~\cite{Allard2009}.

Substituting Eqs.~\eqref{eq:P_kxy} and \eqref{eq:V_kxy} into
Eq.~\eqref{eq:admittance}, we obtain
\begin{equation}
\mathcal{B}
(\mathcal{I} + \mathcal{C}_r)\,
\hat{p}_{\mathrm{i}}(\bm{k}_{xy})
=
\hat{\beta}_0(\bm{k}_{xy})
(-\mathcal{I} + \mathcal{C}_r)\,
\hat{p}_{\mathrm{i}}(\bm{k}_{xy}).
\end{equation}

To construct the normalized acoustic admittance operator \(\mathcal{B}\)
using \((\mathcal{I} + \mathcal{C}_r)^{-1}\),
the invertibility of \((\mathcal{I} + \mathcal{C}_r)\) is required.
This invertibility is guaranteed by the Neumann series expansion in operator theory.
As discussed in the previous section,
\(\mathcal{C}_r\) is a bounded linear integral operator on
\(L^2(\mathcal{D}_k)\),
and its operator norm satisfies
\(\|\mathcal{C}_r\|_{\mathrm{op}} < 1\)
due to energy conservation and the presence of small dissipation.
Therefore,
\[
(\mathcal{I} + \mathcal{C}_r)^{-1}
= \sum_{n=0}^{\infty} (-\mathcal{C}_r)^n,
\]
and invertibility is ensured by norm convergence.

Consequently, the normalized acoustic admittance operator
\(\mathcal{B}\) is given by
\begin{equation}
\mathcal{B}
=
\hat{\beta}_0(\bm{k}_{xy})
(-\mathcal{I} + \mathcal{C}_r)
(\mathcal{I} + \mathcal{C}_r)^{-1}.
\label{eq:B_operator}
\end{equation}

In numerical computations, the continuous wavenumber space is approximated by a
finite-dimensional representation.
By discretizing the wavenumber plane \(\mathcal{D}_k\) into a finite set of points
\(\{\bm{k}_{xy,1}, \bm{k}_{xy,2}, \dots, \bm{k}_{xy,M}\}\),
the matrix representation of the normalized acoustic admittance based on
Eq.~\eqref{eq:B_operator} is obtained as
\begin{equation}
\mathbf{B}
=
\mathbf{B}_0
(-\mathbf{I} + \mathbf{C}_r)
(\mathbf{I} + \mathbf{C}_r)^{-1},
\label{eq:B_matrix}
\end{equation}
where \(\mathbf{I} \in \mathbb{R}^{M \times M}\) is the identity matrix and
\(\mathbf{B}_0 \in \mathbb{R}^{M \times M}\) is a diagonal matrix whose elements correspond
to the characteristic admittance of the medium, defined as
\[
\mathbf{B}_0 :=
\mathrm{diag}\!\left(
\frac{\sqrt{k^2 - (k_{x,1}^2 + k_{y,1}^2)}}{\rho c k},
\;
\frac{\sqrt{k^2 - (k_{x,2}^2 + k_{y,2}^2)}}{\rho c k},
\;
\dots,
\;
\frac{\sqrt{k^2 - (k_{x,M}^2 + k_{y,M}^2)}}{\rho c k}
\right).
\]

Accordingly, for a pressure vector in the wavenumber domain
\(\hat{\mathbf{p}} \in \mathbb{C}^M\),
the corresponding particle velocity vector
\(\hat{\mathbf{v}} \in \mathbb{C}^M\)
is computed as
\[
\hat{\mathbf{v}} = \mathbf{B}\, \hat{\mathbf{p}}.
\]

\subsection{Application of Wavenumber-Domain Admittance to the Boundary Element Method}

The Boundary Element Method (BEM) evaluates sound pressure and particle velocity
on a boundary by discretizing the boundary integral equation derived from the
Kirchhoff--Helmholtz integral theorem, and is widely used for acoustic absorption
and scattering analyses \cite{BurtonMiller1971,Kirkup1998}.

Consider a domain $\Omega$ containing a point source of strength $A$ located at
$\bm{r}_{\mathrm{s}}$.
The sound pressure field $p(\bm{r})$ satisfies the Kirchhoff--Helmholtz integral
equation, where $G$ denotes the free-field Green's function, $k$ the acoustic
wavenumber, and $\rho$ the medium density.
For observation points on the boundary $\Gamma$, the solid-angle coefficient is
$\epsilon=1/2$.
After discretizing $\Gamma$ into $N_{\mathrm{b}}$ boundary elements, the standard
BEM formulation can be written in matrix form as
\begin{equation}
\frac{1}{2}\mathbf{p}
=
A\mathbf{g}_{\mathrm{s}}
+
\mathbf{G}'\mathbf{p}
-
i\omega\rho\,\mathbf{G}\mathbf{v},
\label{eq:bem_standard_half}
\end{equation}
where $\mathbf{p}$ and $\mathbf{v}$ denote the boundary pressure and particle
velocity vectors, respectively.

Conventional BEM assumes a locally reactive boundary by imposing a pointwise
relation $\mathbf{v}=\beta\,\mathbf{p}$.
While this approximation is valid for normal incidence, it fails to capture
direction-dependent behavior at oblique incidence.
To overcome this limitation, the present study introduces a nonlocal boundary
operator defined in the wavenumber domain.

Applying a spatial Fourier transform to the boundary pressure distribution,
the particle velocity on the boundary can be expressed in operator form as
\begin{equation}
\mathbf{v}
=
\widetilde{\mathbf{F}}_{xy}^{\mathrm H}
\mathbf{W}_k
\mathbf{B}
\widetilde{\mathbf{F}}_{xy}\,\mathbf{p}
=
\widehat{\mathbf{B}}\,\mathbf{p},
\label{eq:bem_nonlocal_bc}
\end{equation}
where $\mathbf{B}\in\mathbb{C}^{n_k\times n_k}$ denotes the specific acoustic
admittance matrix in the wavenumber domain and
$\widehat{\mathbf{B}}$ its spatial-domain counterpart.
The matrix $\widetilde{\mathbf{F}}_{xy}\in\mathbb{C}^{n_k\times N_{\mathrm{b}}}$
represents the whitened spatial Fourier transform defined below.
Substituting Eq.~\eqref{eq:bem_nonlocal_bc} into
Eq.~\eqref{eq:bem_standard_half} yields a closed BEM system that can be solved
using standard linear solvers.

The spatial Fourier transform matrix $\mathbf{F}_{xy}$ is constructed as a
discrete approximation of the continuous two-dimensional Fourier transform
defined in Eq.~\eqref{eq:fourier2d_def}.
Using $N_{\mathrm{b}}$ boundary element centers
$\{(x_m,y_m)\}_{m=1}^{N_{\mathrm{b}}}$ on the plane $z=0$ and $n_k$ discrete
in-plane wavenumber vectors $\{\bm{k}_{xy,i}\}_{i=1}^{n_k}$, its elements are
defined as
\begin{equation}
\left[\mathbf{F}_{xy}\right]_{im}
=
\frac{1}{2\pi}
\exp\!\left(-i\,\bm{k}_{xy,i}\cdot\bm{r}_{xy,m}\right),
\label{eq:Fxy_def}
\end{equation}
where $\bm{r}_{xy,m}=(x_m,y_m)$.
The factor $1/(2\pi)$ follows directly from the continuous Fourier transform
convention, whereas the discretization of the integration measure is handled
separately through the quadrature weights introduced in
Subsection~\ref{sec:def_cr}.

Because the wavenumber domain $\mathcal{D}_k$ is discretized, the resulting set
of plane-wave basis functions is not exactly orthogonal under the continuous
energy inner product, and in general
$\mathbf{F}_{xy}^{\mathrm H}\mathbf{W}_k\mathbf{F}_{xy}\neq\mathbf{I}$.
To compensate for this numerical non-orthogonality, a whitening normalization is
applied based on the weighted Gram matrix
\[
\mathbf{H}
=
\mathbf{F}_{xy}^{\mathrm H}\mathbf{W}_k\mathbf{F}_{xy}.
\]
The inverse square root $\mathbf{H}^{-1/2}$ is defined as the Hermitian
positive-definite matrix satisfying
$\mathbf{H}^{-1/2}\mathbf{H}\mathbf{H}^{-1/2}=\mathbf{I}$, and can be computed,
for example, via an eigenvalue decomposition of $\mathbf{H}$.

Using this definition, the normalized spatial Fourier transform matrix is given by
\begin{equation}
\widetilde{\mathbf{F}}_{xy}
=
\mathbf{F}_{xy}\mathbf{H}^{-1/2},
\label{eq:Fxy_whiten}
\end{equation}
which satisfies the weighted orthonormality condition
$\widetilde{\mathbf{F}}_{xy}^{\mathrm H}\mathbf{W}_k\widetilde{\mathbf{F}}_{xy}
=\mathbf{I}$.
Employing $\widetilde{\mathbf{F}}_{xy}$ in
Eq.~\eqref{eq:bem_nonlocal_bc} eliminates artificial coupling between plane-wave
components and enables stable boundary element simulations with
direction-dependent boundary operators beyond locally reactive models.

\section{Numerical Experiment I: Estimation of the Wavenumber-Domain Reflection Coefficient}

\subsection{Simulation Conditions}

To validate the effectiveness of the proposed wavenumber-domain reflection
coefficient estimation method, a numerical simulation was carried out.
Two types of boundaries were considered:
a flat plate exhibiting specular reflection and a periodic slit structure
producing multi-directional scattering.
Both boundaries were modeled as acoustically rigid surfaces (Neumann condition).

\begin{figure}[bht]
\begin{center}
\begin{subfigure}{0.48\textwidth}
  \centering
  \includegraphics[width=\linewidth]{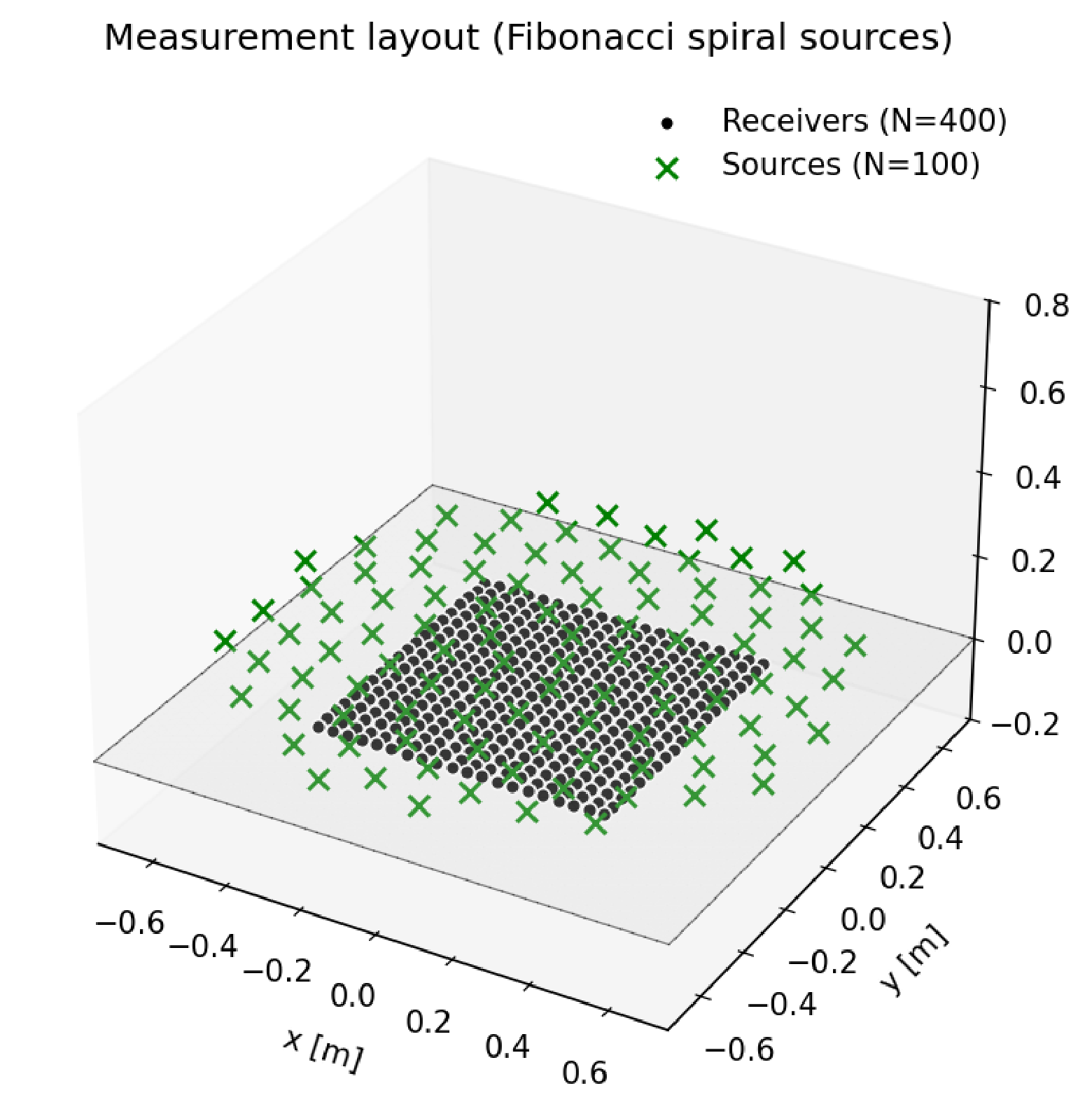}
  \caption{Flat plate boundary condition}
  \label{fig:layout_flat}
\end{subfigure}
\hfill
\begin{subfigure}{0.48\textwidth}
  \centering
  \includegraphics[width=\linewidth]{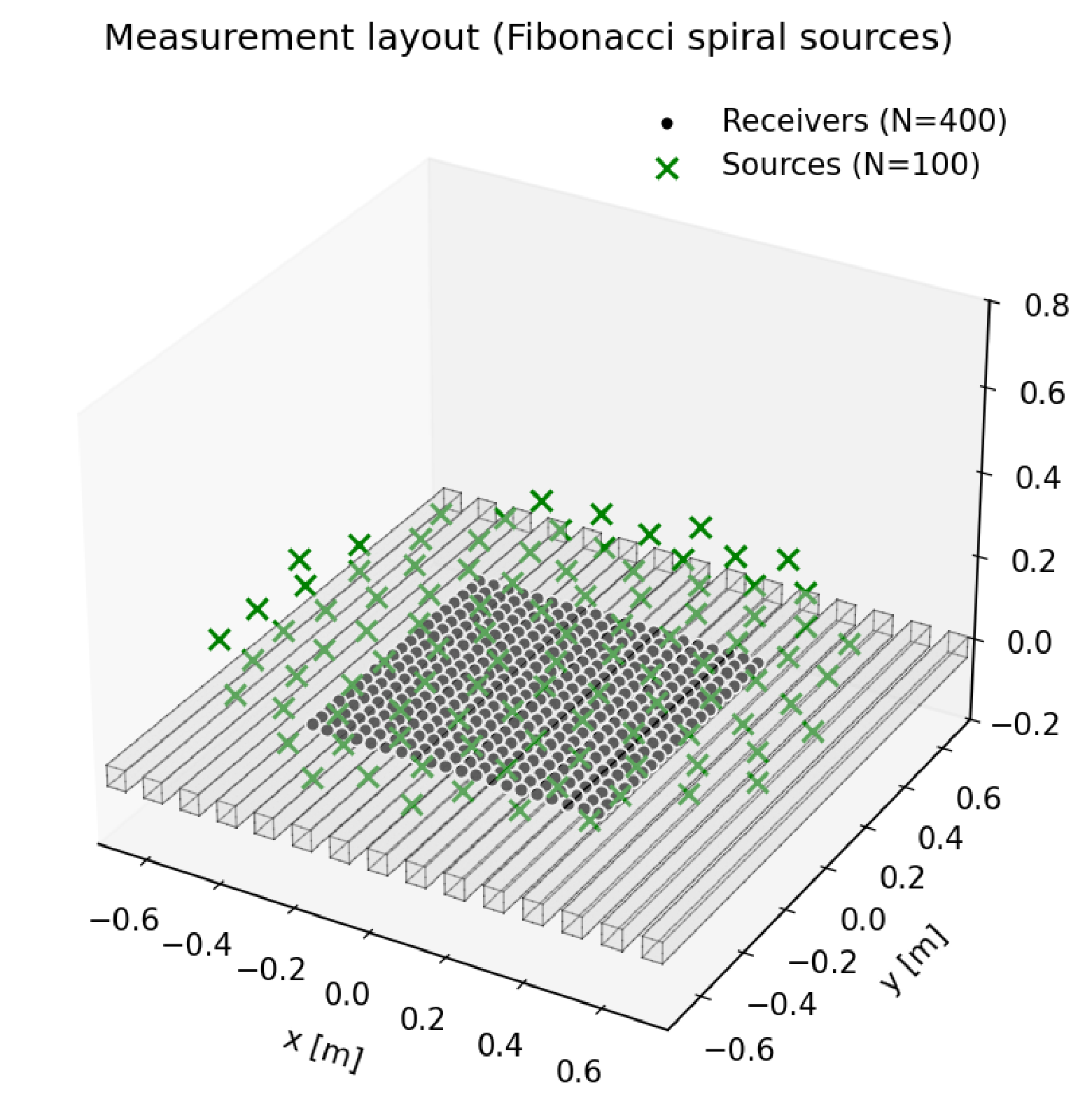}
  \caption{Periodic slit boundary condition}
  \label{fig:layout_slit}
\end{subfigure}
\vspace{-5pt}
\caption{Numerical analysis setup for the two boundary conditions (units in meters).
Omnidirectional point sources (white triangles) are placed on a plane 0.10~m above the boundary surface.
A total of 100 sources are distributed with a fixed radial distance of $\bm{r=0.75}$~m using Fibonacci sampling
to ensure an approximately uniform angular distribution without directional bias.
Receiver points (white circles) are located on a plane 0.01~m above the boundary and arranged in a regular grid
of 400 points with a spacing of 0.04~m.
(a) Flat plate boundary condition; (b) Periodic slit boundary condition.
Both boundaries are modeled as acoustically rigid (Neumann) surfaces.}
\label{fig:measurement_layout}
\end{center}
\vspace{-10pt}
\end{figure}

The sound sources were positioned 0.1~m above the reflecting surface,
and 100 omnidirectional point sources were distributed around this height.
The source height and radial distance ($r = 0.75$~m) were fixed, and the spatial
distribution was determined using Fibonacci sampling to avoid directional bias.
Receiver points were placed 0.01~m above the reflecting surface, arranged in a
regular grid of 400 points with a spacing of 0.04~m.
The source-receiver configuration is illustrated in
Figs.~\ref{fig:layout_flat} (flat plate) and \ref{fig:layout_slit} (slit).

Each source was sequentially excited, and the incident and reflected wavefields
were computed using frequency-domain BEM.
The resulting multi-source/multi-receiver pressure dataset was processed using
$\ell_1$-regularized solver (LASSO\cite{tibshirani1996}) as a practical option to estimate the reflection coefficient
matrix $\mathbf{C}_{\mathrm{r}}$, representing the directional mapping between
incident and reflected wave components in the wavenumber domain.

The slit boundary consisted of rigid bars with a width of 0.05~m arranged
periodically with a pitch of 0.10~m.
The same source and receiver configuration was used for both boundary conditions
to enable a direct comparison.

The excitation frequency was set to 3.4~kHz, such that the acoustic wavelength is
on the same order as the characteristic length scale of the slit geometry.
This choice allows diffraction and multi-directional scattering effects induced
by the periodic structure to be clearly manifested, while remaining within a
frequency range that can be accurately resolved by the numerical discretization.

The speed of sound was set to $c = 343.5$~m/s and the air density to
$\rho = 1.205$~kg/m$^3$.

\subsection{Results}

\begin{figure}[t]
  \centering
  \begin{subfigure}[b]{0.45\linewidth}
    \centering
    \includegraphics[width=\linewidth]{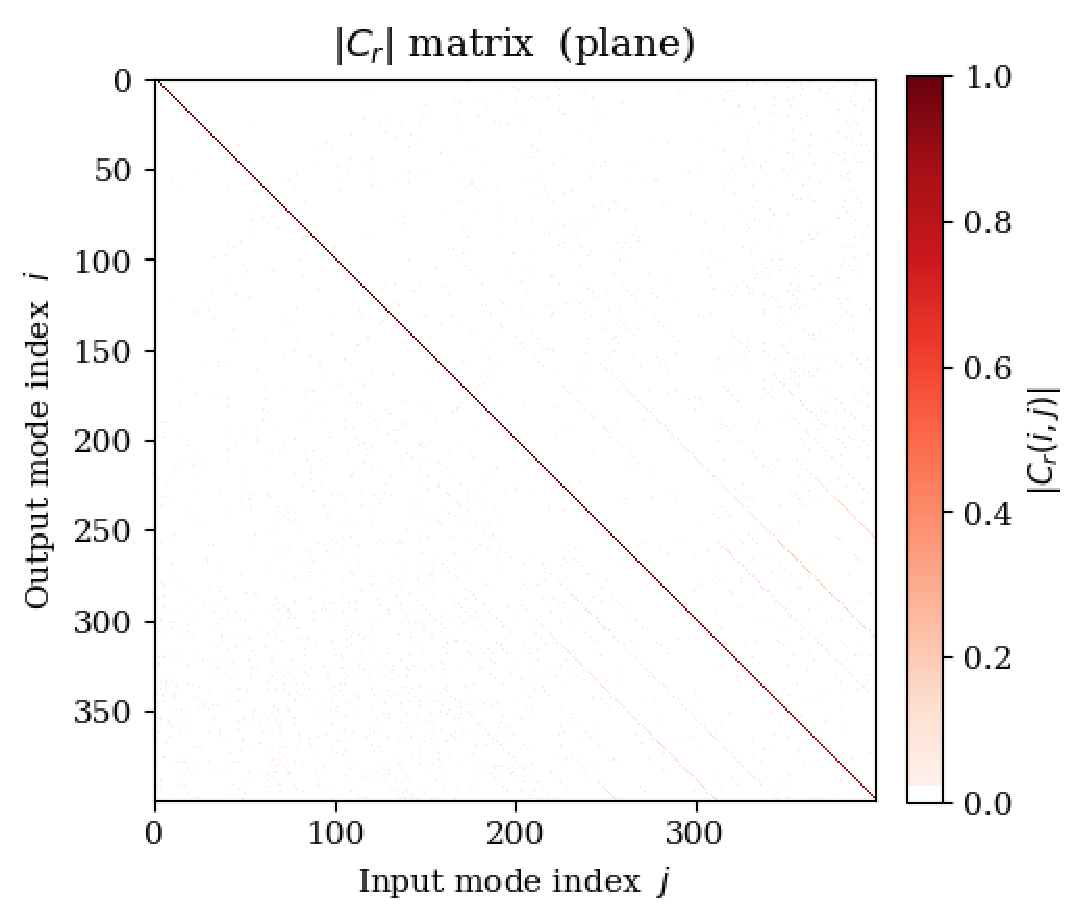}
    \caption{Flat plate}
    \label{fig:Cr_flat}
  \end{subfigure}
  \hfill
  \begin{subfigure}[b]{0.45\linewidth}
    \centering
    \includegraphics[width=\linewidth]{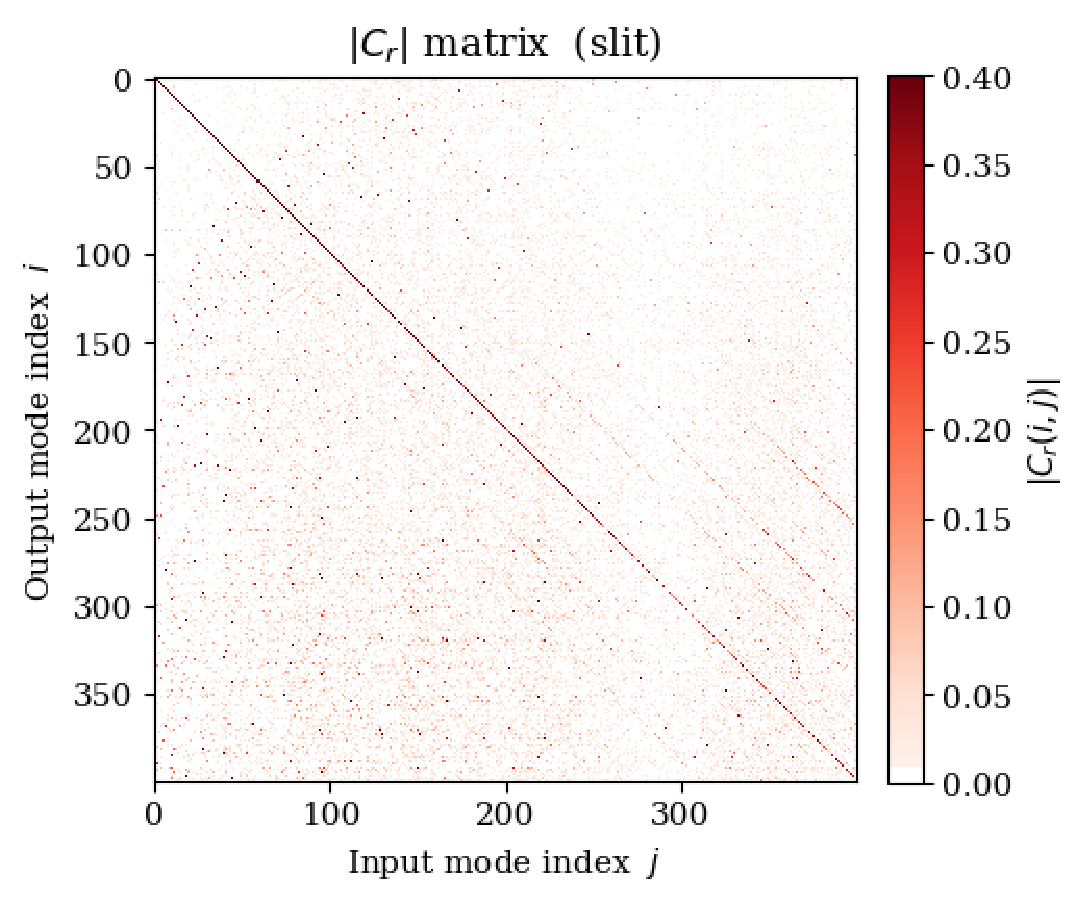}
    \caption{Periodic slit}
    \label{fig:Cr_slit}
  \end{subfigure}

  \caption{
    Estimated wavenumber-domain reflection matrices $\mathbf{C}_\mathrm{r}$.
    For the flat plate, energy is concentrated along the main diagonal,
    indicating dominant specular reflection.
    In contrast, the periodic slit exhibits pronounced off-diagonal components,
    corresponding to multi-directional scattering.
  }
  \label{fig:Cr}
\end{figure}

Figures~\ref{fig:Cr}(a) and \ref{fig:Cr}(b) show the estimated
wavenumber-domain reflection coefficient matrices $\mathbf{C}_\mathrm{r}$
for the flat plate and periodic slit boundaries, respectively.
The horizontal axis represents the incident wavenumber mode index,
and the vertical axis represents the reflected mode index.
For visual clarity, the number of wavenumber bins was set to 400.
For the flat-plate condition shown in Fig.~\ref{fig:Cr}(a),
the matrix energy is strongly concentrated along the main diagonal,
confirming the dominance of specular reflection as theoretically expected.
In contrast, the periodic slit condition shown in Fig.~\ref{fig:Cr}(b)
exhibits pronounced off-diagonal components,
clearly capturing the multi-directional scattering induced by the periodic structure.

\begin{figure}[tb]
  \centering

  \begin{subfigure}{\linewidth}
    \centering

    \begin{minipage}{0.32\linewidth}
      \centering
      \includegraphics[width=\linewidth]{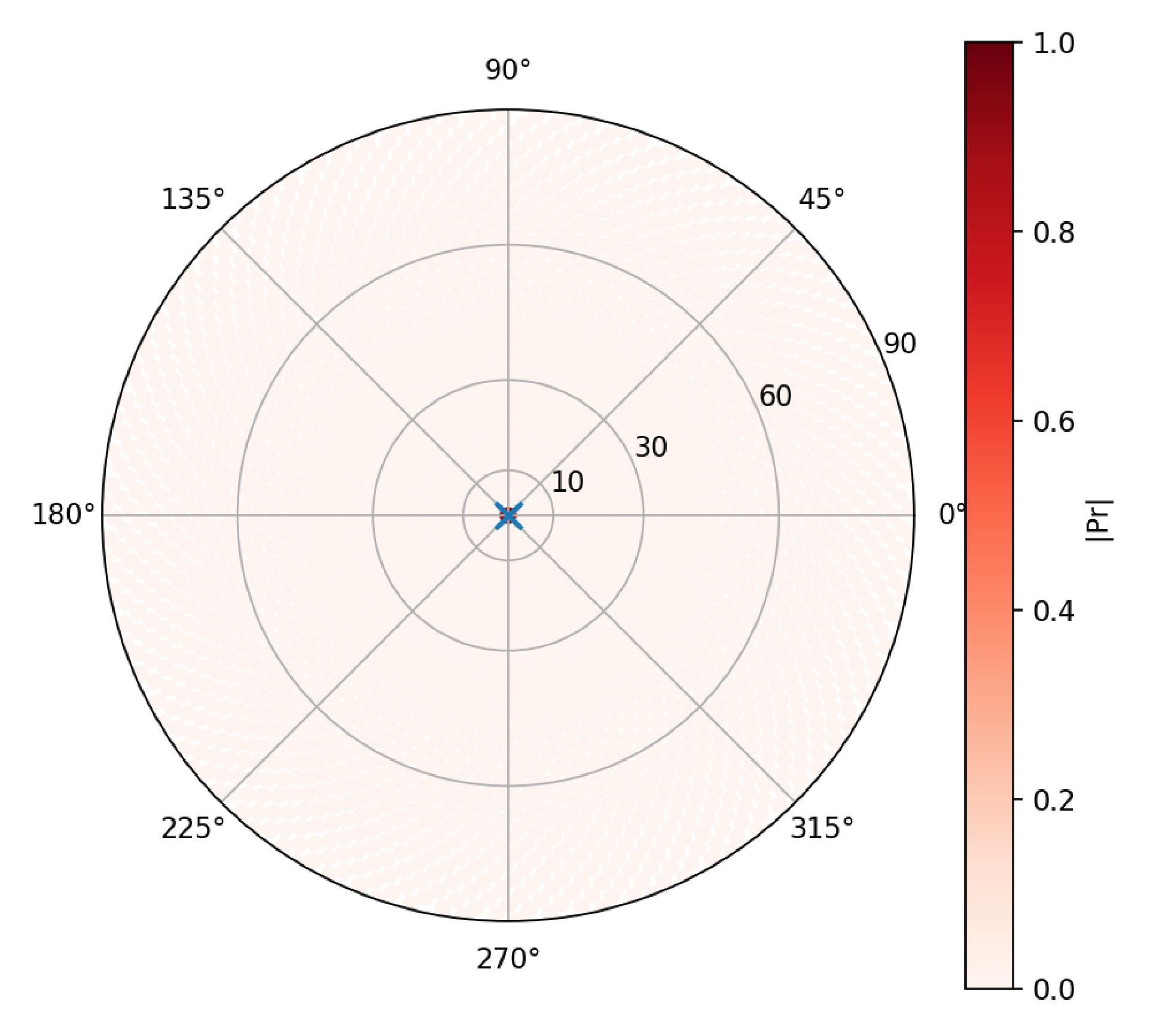}\\
      {\small $\theta_\mathrm{i}=0^\circ$}
    \end{minipage}
    \begin{minipage}{0.32\linewidth}
      \centering
      \includegraphics[width=\linewidth]{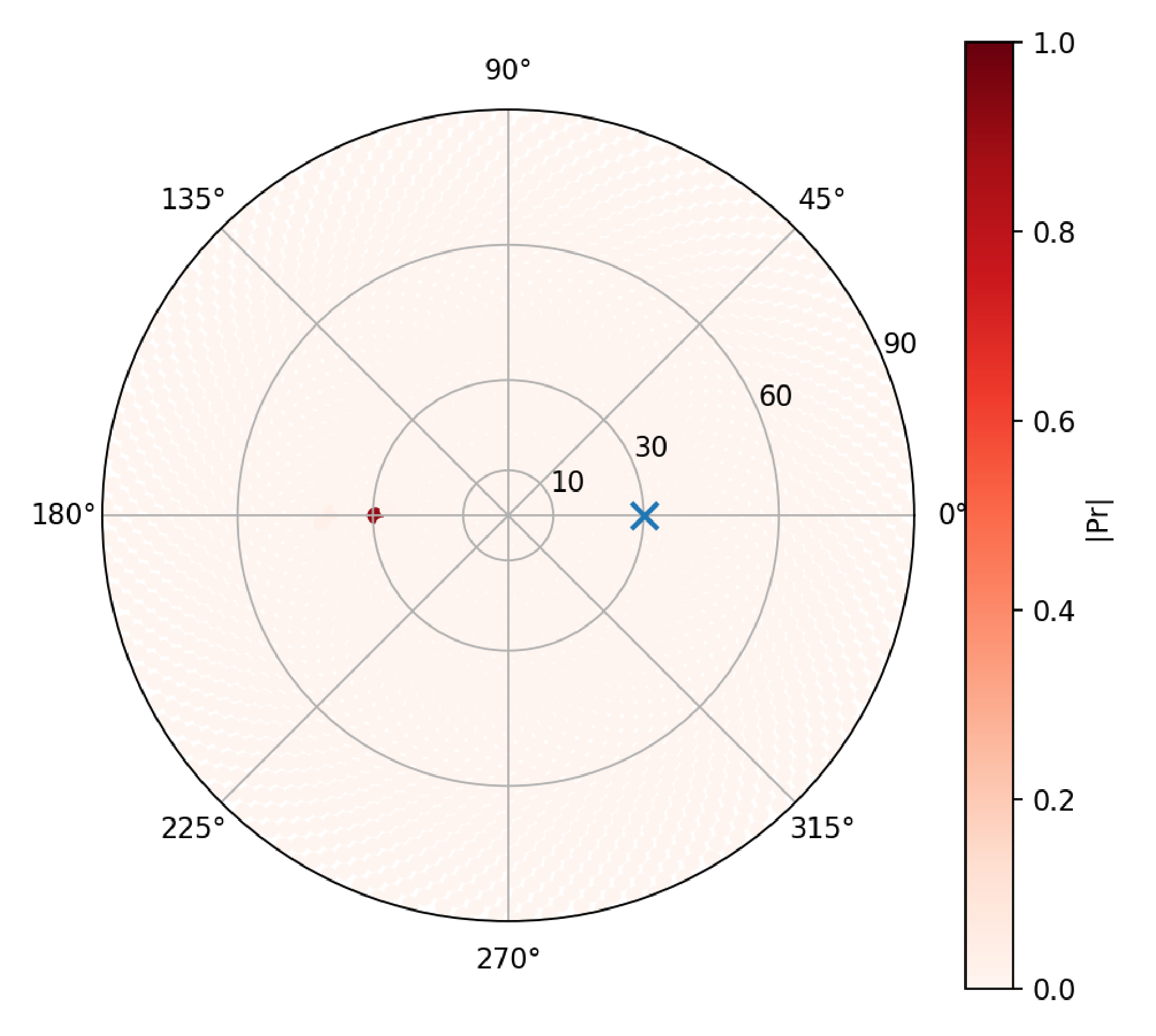}\\
      {\small $\theta_\mathrm{i}=30^\circ$}
    \end{minipage}
    \begin{minipage}{0.32\linewidth}
      \centering
      \includegraphics[width=\linewidth]{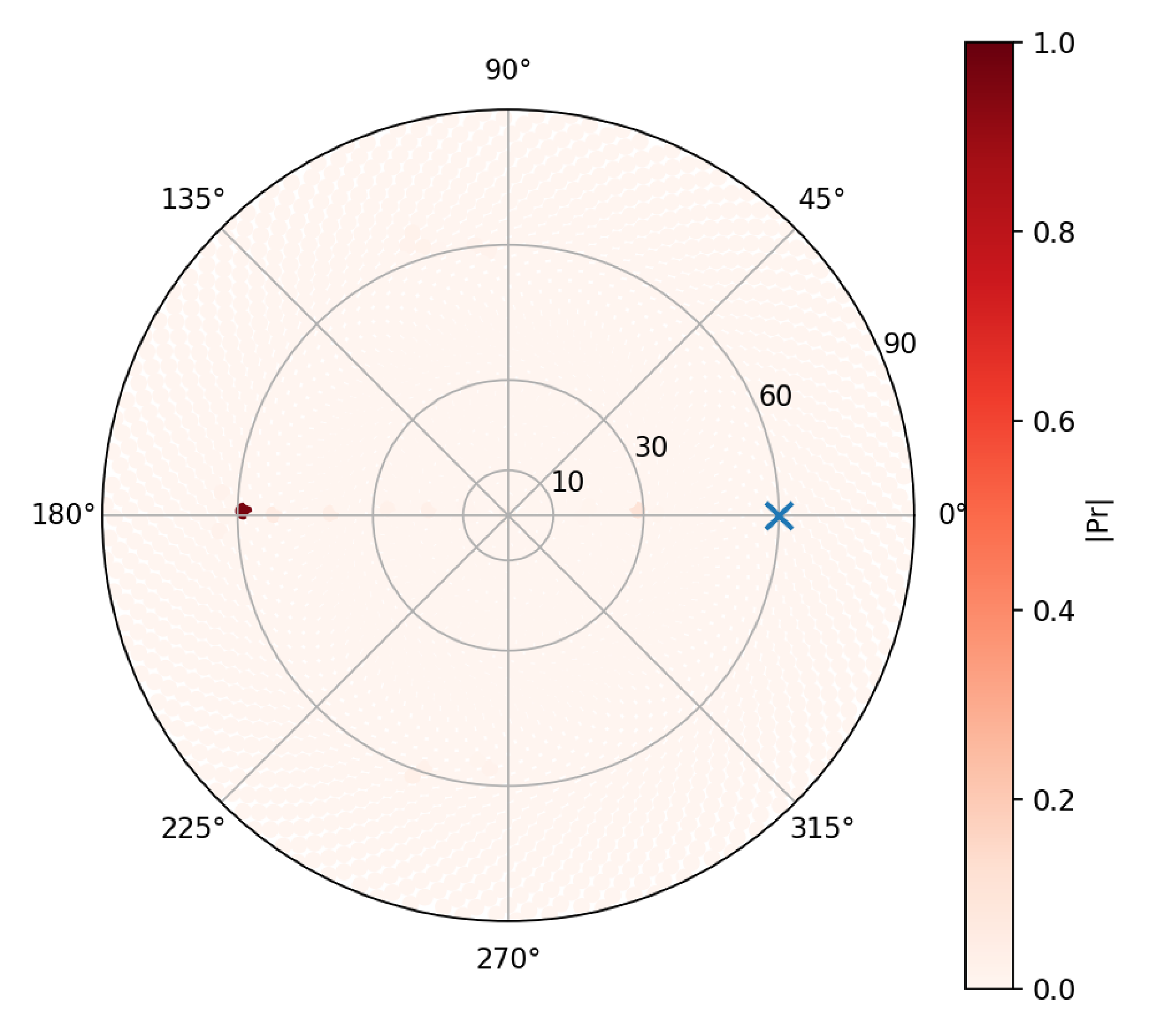}\\
      {\small $\theta_\mathrm{i}=60^\circ$}
    \end{minipage}

    \caption{Flat plate}
    \label{fig:Pr_flat}
  \end{subfigure}

  \vspace{2mm}

  \begin{subfigure}{\linewidth}
    \centering

    \begin{minipage}{0.32\linewidth}
      \centering
      \includegraphics[width=\linewidth]{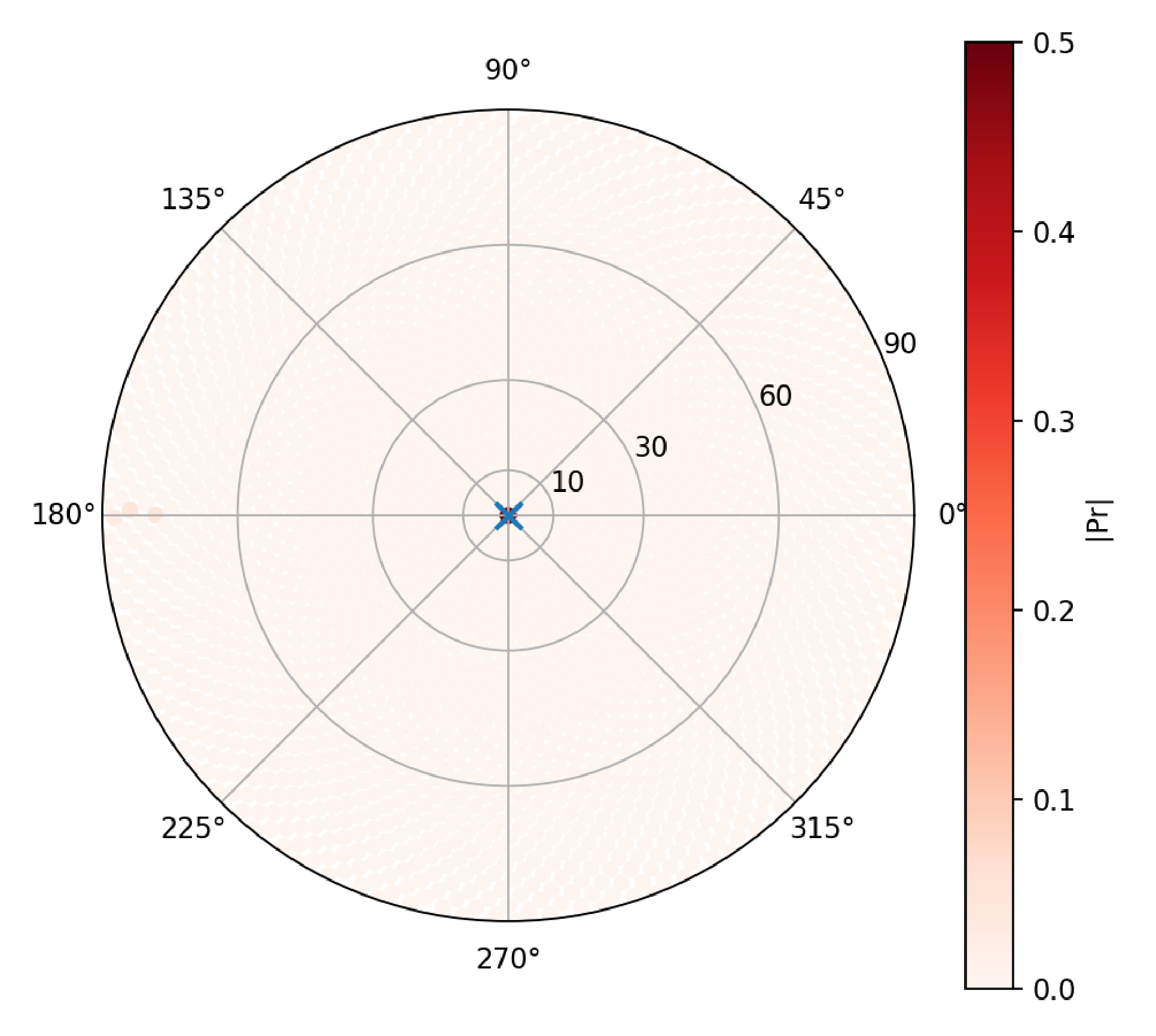}\\
      {\small $\theta_\mathrm{i}=0^\circ$}
    \end{minipage}
    \begin{minipage}{0.32\linewidth}
      \centering
      \includegraphics[width=\linewidth]{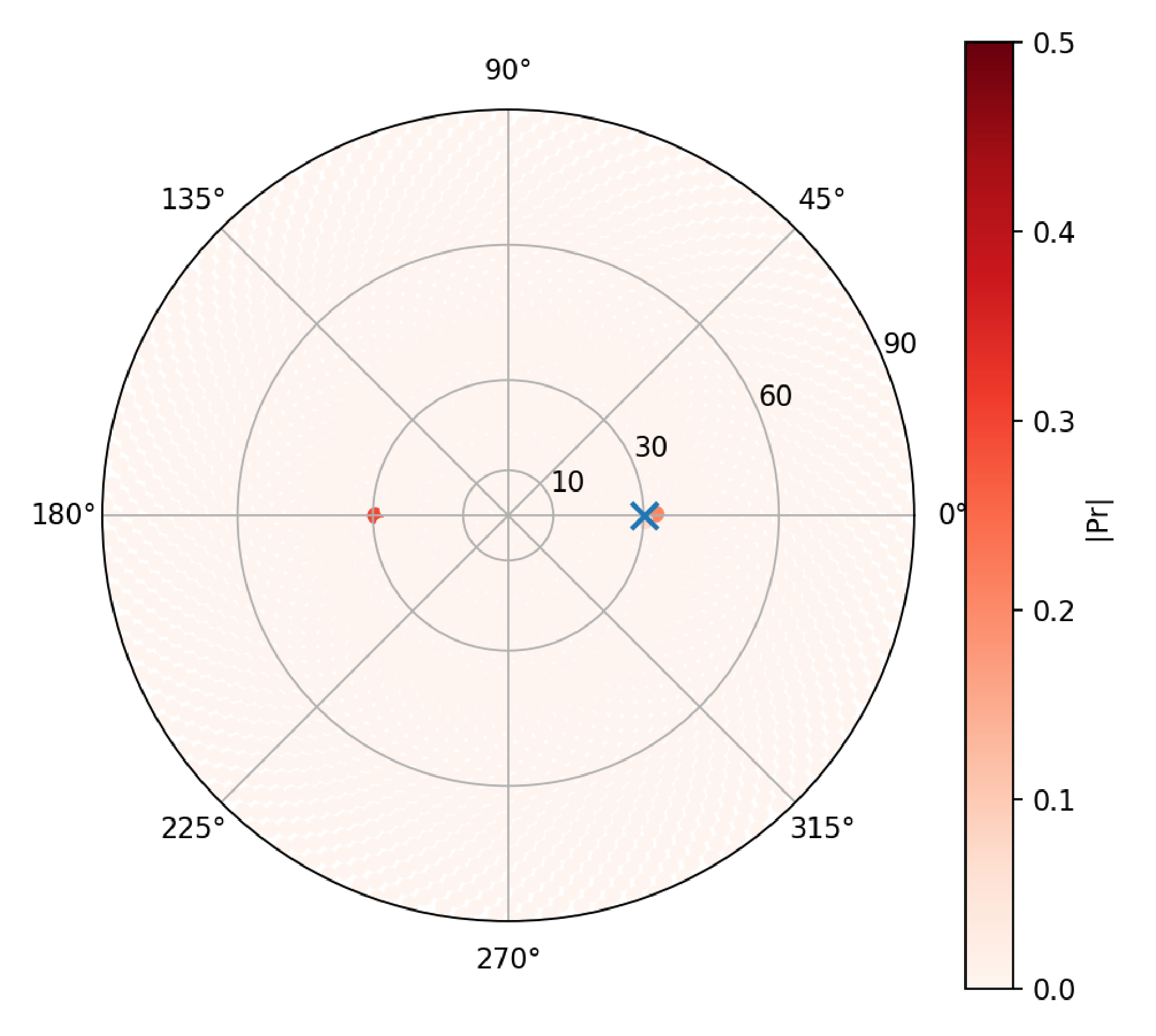}\\
      {\small $\theta_\mathrm{i}=30^\circ$}
    \end{minipage}
    \begin{minipage}{0.32\linewidth}
      \centering
      \includegraphics[width=\linewidth]{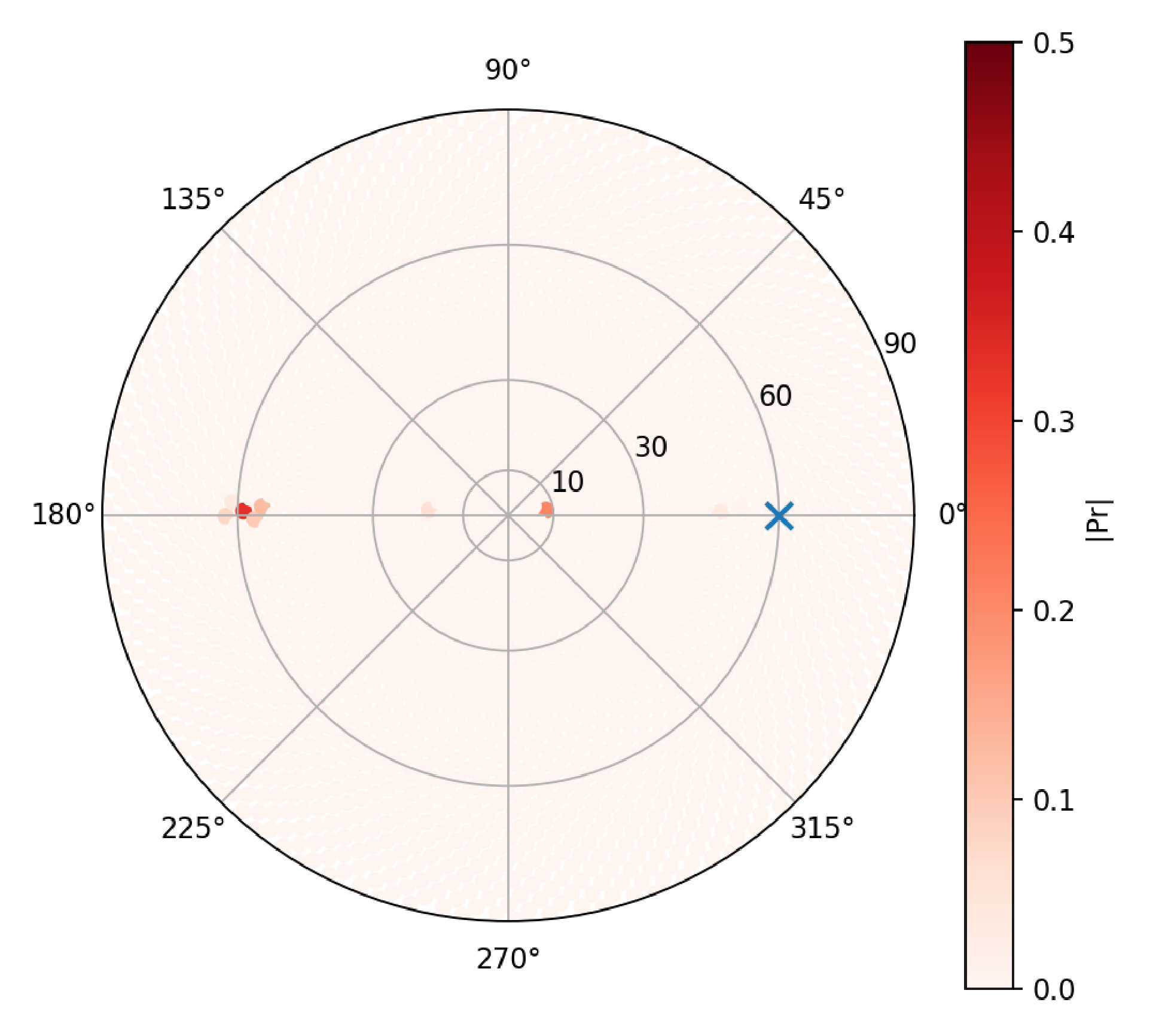}\\
      {\small $\theta_\mathrm{i}=60^\circ$}
    \end{minipage}

    \caption{Periodic slit}
    \label{fig:Pr_slit}
  \end{subfigure}

  \caption{
    Reconstructed reflection directivity maps $|P_\mathrm{r}|$
    computed using the estimated $\mathbf{C}_\mathrm{r}$.
    Blue crosses indicate the incident directions.
  }
  \label{fig:Pr}
\end{figure}

To examine the validity of the estimated $\mathbf{C}_\mathrm{r}$,
reflection directivity maps $|P_\mathrm{r}|$ were reconstructed using the matrices.
Figure~\ref{fig:Pr}(a) presents the reflected pressure distributions
for the flat plate, while Fig.~\ref{fig:Pr}(b) shows the corresponding results
for the periodic slit, under incident angles
$\theta_\mathrm{i} = 0^\circ, 30^\circ,$ and $60^\circ$.
The blue cross marks indicate the incident direction, enabling intuitive visualization
of how acoustic energy is redistributed into specular and scattered directions.
The incident-wave vector $\mathbf{\hat{p}}_\mathrm{i}$ used for the reconstruction
was defined as a one-hot vector, in which only the component corresponding
to the incident direction was set to unity.
This formulation allows direct evaluation of how accurately
the estimated wavenumber-domain reflection matrix $\mathbf{C}_\mathrm{r}$
reproduces the directional reflection behavior.
To ensure sufficient angular resolution,
a Fibonacci lattice with 2401 wavenumber bins was employed for the reconstruction.
Owing to the uniform angular sampling, both specular and scattered components
are smoothly visualized over the entire angular range.

These results demonstrate that the proposed method robustly and accurately estimates
the wavenumber-domain reflection matrix $\mathbf{C}_\mathrm{r}$ for a wide variety of boundary conditions,
ranging from purely specular surfaces to strongly scattering periodic structures.
Furthermore, the reconstructed reflection maps confirm that directional reflection behavior
can be reproduced with high fidelity using the estimated operator.

\section{Numerical Experiment II: Sound-Field Simulation Using \\ the Wavenumber-Domain Acoustic Admittance}
\subsection{Simulation Conditions}

In this section, the wavenumber-domain acoustic reflection coefficients estimated in the previous chapter
are converted into their admittance representation and incorporated into the Boundary Element Method (BEM)
to evaluate sound-field reproduction performance.
Two BEM approaches are compared:

(i) a conventional BEM in which the flat and slit reflectors are explicitly meshed in detail, and  
(ii) the proposed BEM, in which only the flat surface is meshed and the wavenumber-domain admittance is imposed as a boundary condition.

The reflector surface was positioned such that its upper boundary corresponds to $z=0$~m.
A monopole source was placed at $(x,y,z)=(0~\mathrm{m},0~\mathrm{m},0.4~\mathrm{m})$ to excite a steady-state field at 3400~Hz.

In the conventional BEM, discretizing the reflector geometry with a mesh size of approximately 30~mm
resulted in 5,830 boundary elements for the flat plate and 15,180 elements for the slit reflector.
In contrast, the proposed BEM employs only the 5,830-element flat-surface mesh,
using the same mesh even for the slit condition while imposing the wavenumber-domain admittance
estimated in the previous chapter.

Sound pressure was evaluated on two orthogonal cross sections ($x=0$ and $y=0$)
over a region of $1.5~\mathrm{m} \times 1.0~\mathrm{m}$
with observation points placed at 0.05~m intervals.
The speed of sound was set to $c=343.5$~m/s.

\subsection{Results}

\begin{figure}[t!]
  \centering

  \begin{minipage}[b]{0.49\linewidth}
    \includegraphics[keepaspectratio, width=\linewidth]{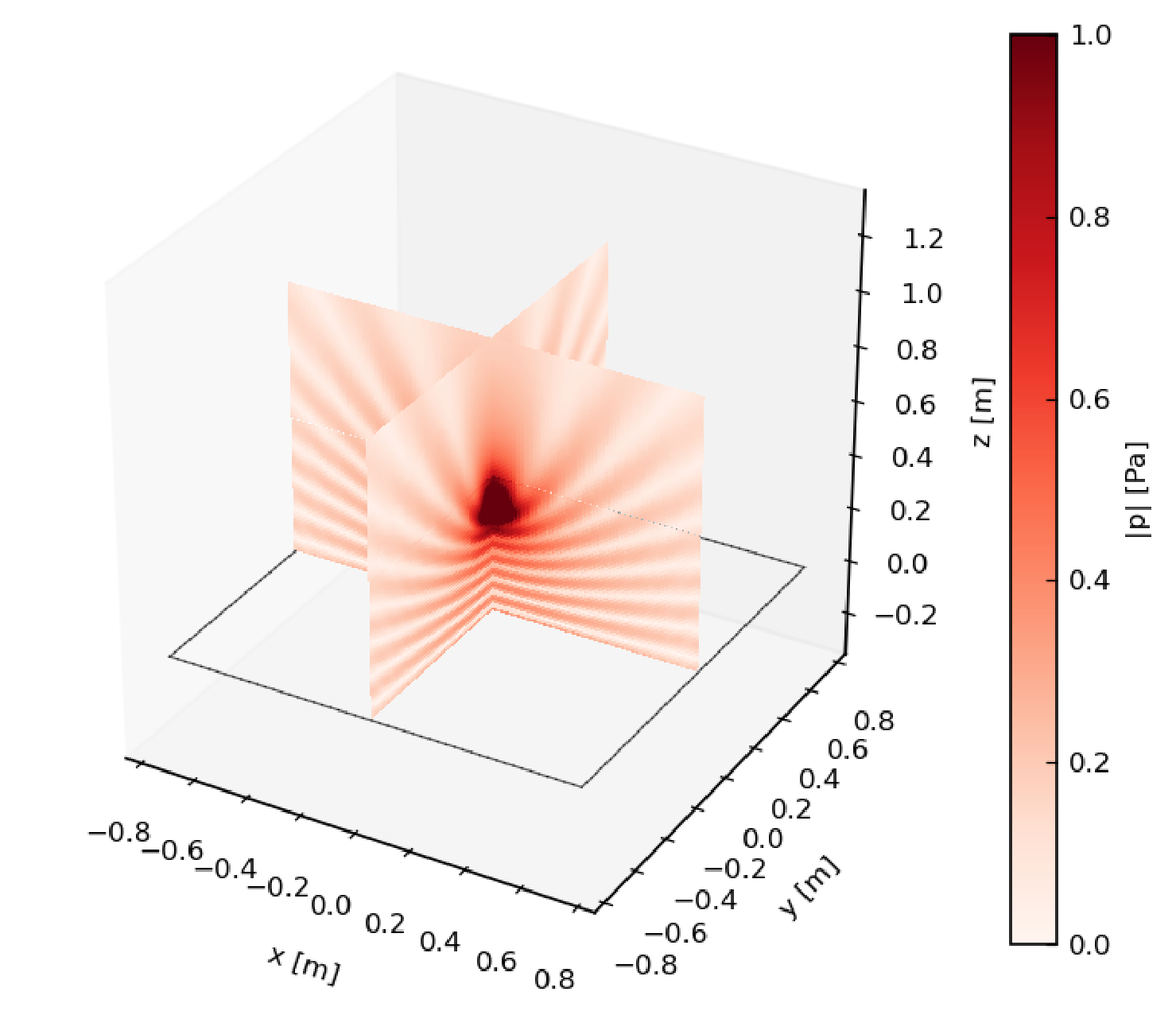}
    \subcaption{Conventional BEM (Full geometry mesh, flat plate)}
    \label{fig:bem_mesh_flat}
  \end{minipage}
  \hfill
  \begin{minipage}[b]{0.49\linewidth}
    \includegraphics[keepaspectratio, width=\linewidth]{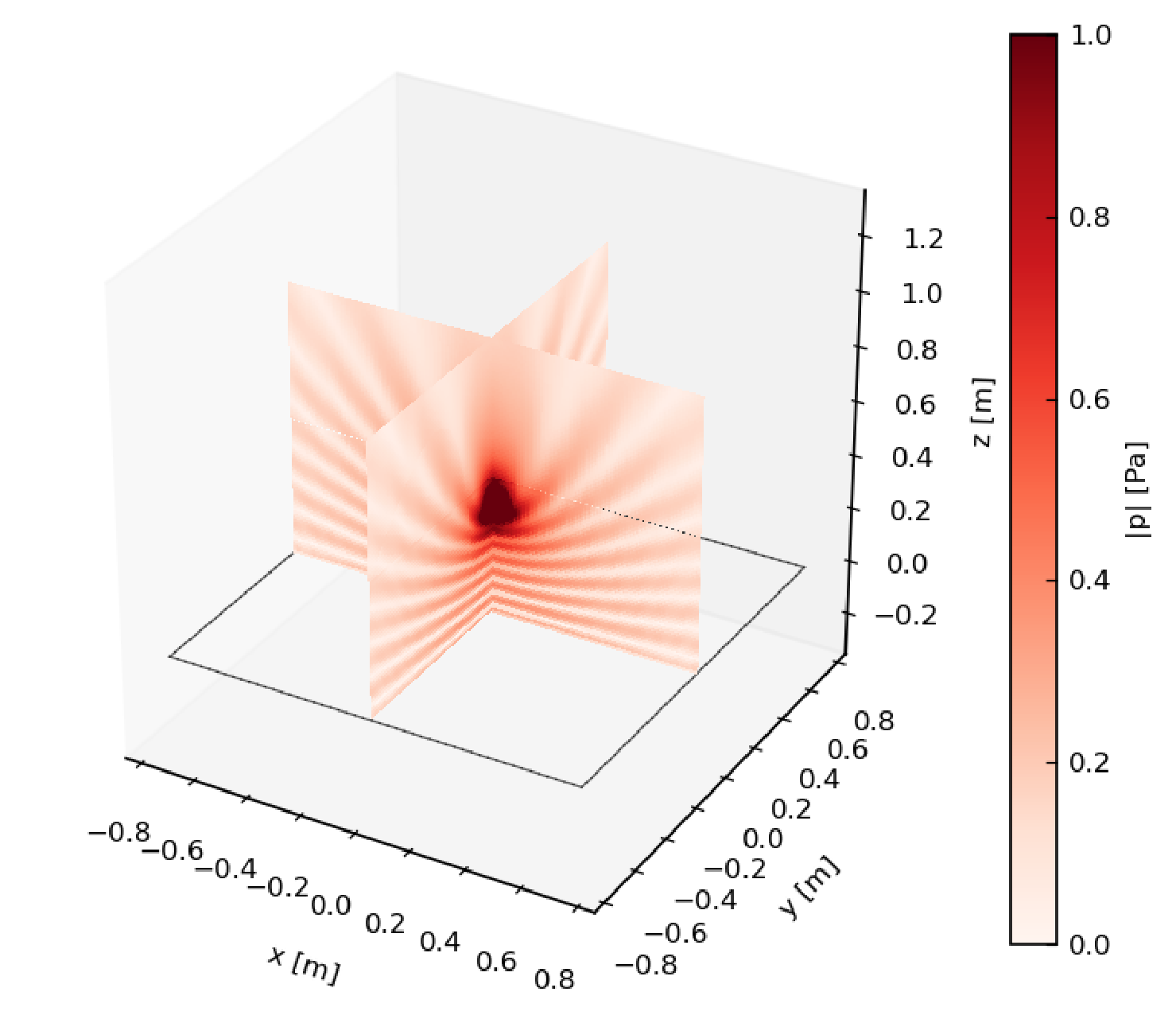}
    \subcaption{Proposed BEM (Surface-only mesh, flat plate)}
    \label{fig:bem_surface_flat}
  \end{minipage}

  \vspace{10pt}

  \begin{minipage}[b]{0.49\linewidth}
    \includegraphics[keepaspectratio, width=\linewidth]{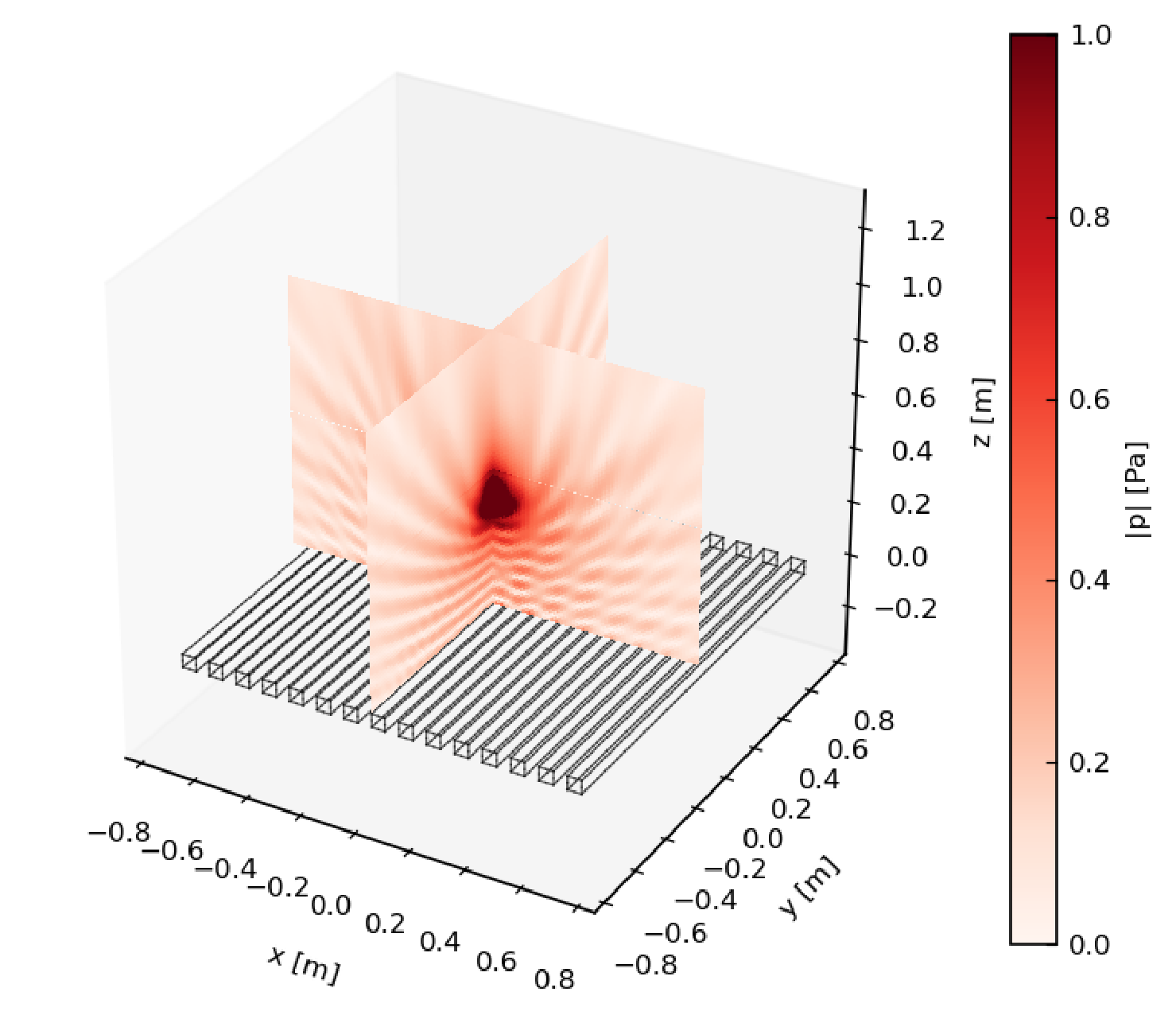}
    \subcaption{Conventional BEM (Full geometry mesh, slit)}
    \label{fig:bem_mesh_slit}
  \end{minipage}
  \hfill
  \begin{minipage}[b]{0.49\linewidth}
    \includegraphics[keepaspectratio, width=\linewidth]{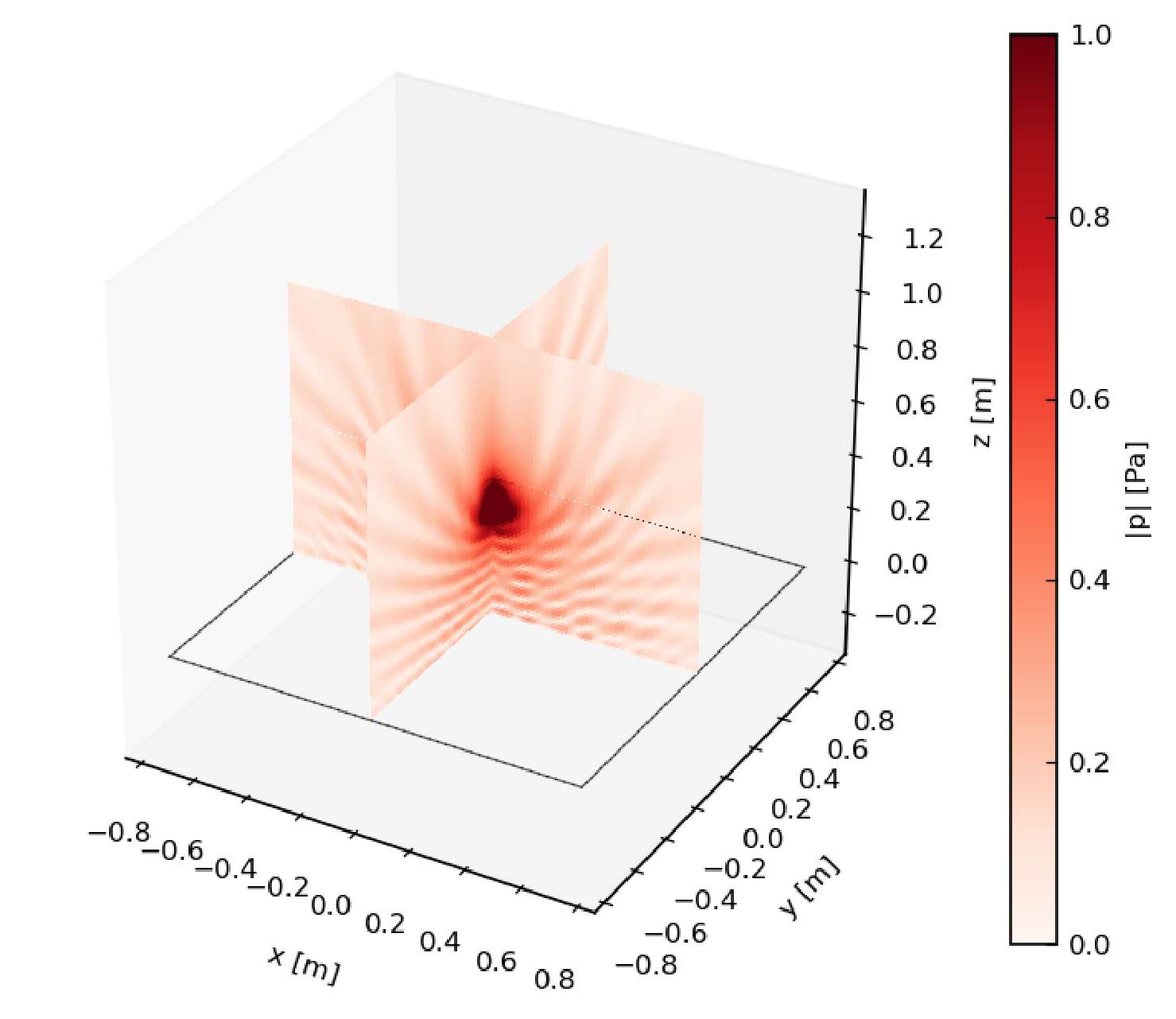}
    \subcaption{Proposed BEM (Surface-only mesh, slit)}
    \label{fig:bem_surface_slit}
  \end{minipage}

  \caption{
    Comparison of total sound pressure magnitude distributions obtained by the conventional geometry-based BEM and the proposed surface-based BEM. Two boundary conditions are considered: a flat plate and a slit structure. The conventional BEM employs 5,830 boundary elements for the flat plate and 15,180 elements for the slit geometry, whereas the proposed method reproduces both cases using only the flat-surface mesh with 5,830 elements, achieving acoustic fields comparable to those predicted by the full-geometry BEM.
  }
  \label{fig:comparison_abs}
\end{figure}

Figure~\ref{fig:comparison_abs} compares the magnitude distributions of the total acoustic field
(incident + reflected) obtained using the conventional BEM and the proposed BEM
for both the flat and slit boundary conditions.

For the flat-plate condition, the sound fields produced by the two methods agree almost perfectly,
indicating that specular reflection is reproduced with high accuracy.
Under the slit condition, a highly complex scattering field is formed due to diffraction and interference
caused by the openings.  
The proposed BEM successfully reconstructs these features with accuracy comparable to that of the conventional BEM.

A quantitative comparison was performed by computing the cosine similarity and mean squared error (MSE)
relative to the conventional BEM results.
For the flat-plate case, the similarity was 0.982 with an MSE of $5.73 \times 10^{-3}$.
For the slit case, the similarity was 0.981 with an MSE of $5.74 \times 10^{-3}$.
Both results indicate excellent agreement.

A particularly noteworthy finding is that the proposed BEM achieves this accuracy
with far fewer boundary elements.
Whereas the conventional BEM required 5,830 elements for the flat plate
and 15,180 elements for the slit geometry,
the proposed method reproduces slit-induced diffraction and scattering
using only the simple flat-surface mesh with 5,830 elements,
which is identical to that used for the flat-plate case.

This demonstrates the effectiveness of externally prescribing scattering behavior
through the estimated wavenumber-domain admittance, rather than explicitly meshing fine geometric details.

These results confirm that the proposed BEM based on wavenumber-domain admittance
can reproduce sound fields with accuracy comparable to conventional full-geometry BEM
while reducing the number of elements by more than half.

\section{Conclusion}

This study introduced a wavenumber-domain acoustic reflection coefficient
$\mathbf{C}_{\mathrm{r}}$ formulated as a linear operator that represents
direction-dependent reflection and scattering behavior at an acoustic boundary.
By focusing exclusively on the relationship between incident and reflected wave
components at the boundary, the proposed framework provides a macroscopic
description of reflection phenomena without explicitly modeling the internal
structure of the material.

An estimation framework was developed in which the directional spectra of
incident and reflected waves are extracted from multi-source, multi-receiver
sound-pressure data via a spatial Fourier transform, and the reflection matrix
$\mathbf{C}_{\mathrm{r}}$ is directly estimated from these spectra.
Although sparsity-promoting solvers can be advantageous for structured or
periodic boundaries, the formulation itself is independent of the specific
estimation algorithm.

Once estimated, $\mathbf{C}_{\mathrm{r}}$ can be converted into a
wavenumber-domain acoustic admittance and incorporated into conventional BEM
formulations as a nonlocal boundary condition.
This enables sound-field analysis without explicitly meshing fine geometric
features, allowing complex reflection and scattering characteristics to be
imposed directly at the boundary.

Three-dimensional BEM validations demonstrated that the proposed method
accurately reproduces both specular reflection from a flat plate and
multi-directional scattering from a periodic slit structure.
The resulting sound fields showed excellent agreement with those obtained using
geometry-resolved BEM, while significantly reducing modeling complexity.

Unlike conventional scalar reflection models defined independently for each
incident angle, the proposed matrix-valued formulation captures both directional
dependence and scattering-induced coupling in a unified framework.
Future work will focus on experimental estimation for real materials and
extensions to more complex acoustic boundaries.

\section*{Acknowledgments}
This work was supported by JSPS KAKENHI Grant Number JP24K03222.

\end{document}